\documentclass[amsfonts, amsmath, amssymb, aps, pra, twocolumn, a4paper, superscriptaddress, floatfix]{revtex4-2}
\usepackage[utf8]{inputenc}
\usepackage{graphicx}
\usepackage{braket}
\usepackage{xcolor}
\usepackage{physics}
\usepackage{mathtools}
\usepackage{qcircuit}
\usepackage{amsmath}
\usepackage{amssymb}
\usepackage{supertabular}
\usepackage{subfigure}
\usepackage{soul}
\usepackage{multirow}
\usepackage[english]{babel}
\usepackage{comment}
\usepackage[T1]{fontenc}
\usepackage{natbib}
\usepackage[ruled,vlined,linesnumbered]{algorithm2e}
\usepackage[export]{adjustbox}
\usepackage[colorlinks=true,citecolor=blue,linkcolor=red,urlcolor=black]{hyperref}

\newcommand{\beq}{\begin{equation}}
\newcommand{\eeq}{\end{equation}}

\newcommand{\lnorm}[2][p]{\norm{#2}} 

\newcommand{\ltnorm}[2][2]{\norm{#2}_{\ell_{#1}}} 

\newcommand{\lonorm}[2][1]{\norm{#2}_{{#1}}} 

\begin{document}

\title{Efficient quantum interpolation of natural data}

\author{Sergi Ramos-Calderer}
\affiliation{Quantum Research Centre, Technology Innovation Institute, Abu Dhabi, UAE.}
\affiliation{Departament de F\'isica Qu\`antica i Astrof\'isica and Institut de Ci\`encies del Cosmos (ICCUB), Universitat de Barcelona, Mart\'i i Franqu\`es 1, 08028 Barcelona, Spain.}

\begin{abstract}
We present efficient methods to interpolate data with a quantum computer that complement uploading techniques and quantum post-processing. The quantum algorithms are supported by the efficient Quantum Fourier Transform (QFT) and classical signal and imaging processing techniques, and open the door of quantum advantage to relevant families of data. We showcase a QFT interpolation method, a Quantum Cosine Transform (QCT) interpolation geared towards natural data, and we improve upon them by utilizing a quantum circuit's capabilities of processing data in superposition. A novel circuit for the QCT is presented. We demonstrate the methods on probability distributions and quantum encoded images, and discuss the precision of the resulting interpolations.
\end{abstract}
\date{\today}

\maketitle

\section{Introduction}

Uploading a large amount of classical data into a quantum state remains a bottleneck for quantum computation applications. Quantum states indeed offer a large Hilbert space to encode classical data, but uploading one element at a time makes a quantum approach inefficient, no matter its promises regarding processing power on the uploaded state. New strategies are needed to overcome the initial threshold of uploading data into a quantum state.

To this end, we present different methods that interpolate smooth probability distributions and natural data over a larger space that can alleviate the data uploading effort, dramatically in some cases. This methods build on classical resampling techniques that employ the Fourier Transform to interpolate band-limited signals. The original amplitude encoded distribution is first Fourier transformed, then complemented with vanishing high frequencies and, finally, an inverse Fourier transform over the larger space delivers the interpolated probability distribution. To be precise, the relevant fact that makes this approach useful when interpolating classical distributions in a quantum computer is that the Quantum Fourier Transform (QFT) is efficient, that is, it only needs a polynomial number of operations as a function of the number of qubits involved.

The QFT interpolation scheme was adapted with success in the context of Tensor Networks in Ref. \cite{garcia2021quantum}, and introduced to enhance the result of solutions of partial differential equations in a quantum computer in Ref. \cite{garcia2022quantum}, where the accuracy of the interpolation is studied. We expand upon this concept with more diverse efficient quantum transformations and further utilize superposition for more efficient interpolation algorithms.

These interpolation methods can also be understood as a discrete-to-discrete instance of the Nyquist-Shannon sampling theorem. The sampling theorem states that all the information of a signal with finite band-width can be captured by samples obtained at a finite rate, known as Nyquist rate. We incorporate these ideas when discussing the accuracy of the proposed resampling techniques. The use of Fourier-like transforms is ubiquitous when reconstructing continuous signals from samples. The QFT can introduce efficient counterparts to techniques used in classical signal processing, and we present some examples in the context of data interpolation and image resampling.

Classical image processing of natural data has been drawn towards the Discrete Cosine Transforms (DCT) \cite{ahmed1974discrete} over the QFT, since the real-to-real mapping  yields better results for common types of images. This transformation is also the basis of the widely used JPEG compression scheme \cite{wallace1992jpeg}. We will incorporate these ideas into the quantum regime in order to improve upon QFT interpolation for natural data.

In Sec. \ref{sec:qft} we re-state the QFT interpolation scheme for probability distributions and study the accuracy of the method when specifically applied to quantum states encoding probability distributions. In Sec. \ref{sec:qct} we detail a novel quantum DCT circuit and utilize it in a Quantum Cosine Transform (QCT) interpolation method for natural images. We discuss the power of processing data in superposition that quantum computers allow. In Sec \ref{sec:qjpeg} we further exploit quantum superposition for a more efficient JPEG-inspired quantum interpolation procedure. We conclude with a discussion on the practical applicability of the method and future directions.


\section{Efficient QFT interpolation for smooth distributions}\label{sec:qft}

We showcase here the basic algorithm for interpolating a quantum-uploaded distribution to higher precision. The algorithm is detailed, then discussed in the context of enhancing quantum data uploading techniques. Afterwards, we study the accuracy of the interpolated distribution.

\begin{figure*}[t!]
    \centering
    \includegraphics[width=\linewidth]{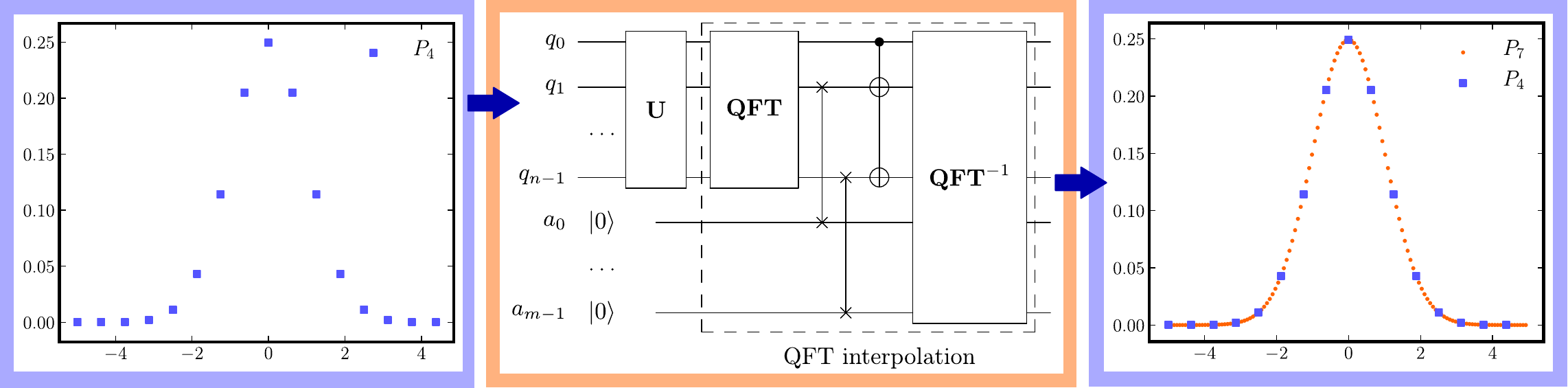}
    \put(-486,106){\textbf{a})}
    \put(-344,106){\textbf{b})}
    \put(-131,106){\textbf{c})}
    \caption{Full process of interpolating a probability distribution $P_n$ using the efficient QFT. a) Shows an example of a Gaussian distribution $P_4$ encoded in $4$ qubits. b) Illustrates the quantum circuit that performs QFT interpolation. Here, $U$ denotes a unitary transformation that uploads the probability distribution $P_n$ into the $q$ quantum register. After an initial QFT of register $q$, a clean ancilla register $a$ is swapped to the first $m$ positions after $q_0$, an operation that can be done virtually by keeping track of qubit position. After applying a CNOT gate to all ancilla qubits controlled by $q_0$, an inverse QFT is applied to the whole system in order to recover the interpolated probability distribution $P_{n+m}$ in the larger space. c) Showcases the interpolation of the initial probability distribution $P_4$ using $n=4$ qubits (blue squares) and the final result $P_7$ over the full $n+m=7$ qubit register (orange dots), normalized to match $P_4$.}
    \label{fig:fig1}
\end{figure*}

Given a quantum register $q$, with individual qubits $q_i$ from $0$ to $n-1$, where a distribution $P$ has been amplitude encoded into a $2^n$ discrete version $P_n$, QFT interpolation proceeds as follows. Apply the QFT to the quantum register $q$. In Fourier space, the high-frequency modes of smooth distributions will be suppressed, or zero in the case of band-limited signals. Therefore, one can artificially pad the high frequency components with quantum states at zero amplitude and not fundamentally alter the original signal. This can be achieved on a quantum circuit by adding an ancillary qubit register $a$, with qubits $a_j$ from $0$ to $m-1$, between original qubits $q_0$ and $q_1$ (the first and second most significant qubits) and then applying CNOT gates, controlled by $q_0$, targeting all qubits in the ancilla register $a$. Finally, an inverse QFT is applied to the entire quantum system. 

The outcome of this circuit is the interpolation from the initial $2^n$ distribution to a larger $2^{n+m}$ space. Fig. \ref{fig:fig1} a) and c) showcase the interpolation of the original $4$ qubit distribution to a larger $7$ qubit space, and implemented using the quantum simulation library \texttt{Qibo} \cite{efthymiou2021qibo, stavros_efthymiou_2022_6080546}. In Fig. \ref{fig:fig1} b) we present the QFT interpolation algorithm on a quantum circuit.

The QFT interpolation algorithm for probability distributions is efficient. That being said, the cost of uploading the initial distribution into the small space needs to be taken into account. There are multiple proposals for uploading a probability distribution into a quantum register, some exact \cite{grover2002creating, shende2006synthesis, kitaev2008wavefunction, plesch2011stateprep, holmes2020efficient, rattew2021efficient} and some training a quantum generator circuit \cite{lloyd2018quantum, dallaire2018quantum, zoufal2019quantum}. If the cost of the initial uploading is already prohibitive this algorithm will only provide a marginal advantage. Still, this technique opens the door for uploading methods that are effective for a small number of qubits that can later be enhanced via QFT interpolation. 

We showcase one such case in App. \ref{appendix:unary}, where we use the QFT interpolation algorithm to enhance a probability distribution uploaded in the unary basis \cite{ramos2021quantum}. This uploading technique trades practical scalability for a device-friendly uploading method that is effective for a small number of amplitudes. After the original distribution is uploaded in unary using $2^n$ qubits, it is transformed into binary to $n$ qubits, where the full available $2^{2^n}$ Hilbert space is reclaimed via QFT interpolation.

In the following, we aim to establish the accuracy of the interpolated distribution when compared to an ideal uploading of the underlying distribution to the larger space. We study the effect of this procedure when applied to the amplitudes of a quantum state.

We will bound the operational distinguishability between the interpolated quantum state and the ideal one by introducing their \textit{trace-distance},
\begin{equation}
    \text{dist}_{\text{Tr}}(\ket{\psi},\ket{\phi})=\frac{1}{2}\lonorm{\ket{\psi}\bra{\psi}-\ket{\phi}\bra{\phi}},
\end{equation}
where $\lonorm{A}=\text{Tr}\,\sqrt{A^\dagger A}$. However, we first proceed by analyzing the $\ell_2$ norm difference between them. The reason for this choice is twofold. Quantum states are normalized under their $\ell_2$ norm, $\ltnorm{\ket{\psi}}=1$ where $\ket{\psi}$ is any pure quantum state, and the QFT preserves this norm, that is $\ltnorm{\ket{\psi}-\ket{\phi}}=\ltnorm{\ket{\Psi}-\ket{\Phi}}$ with $\ket{\Psi}=QFT\ket{\psi}$ and $\ket{\Phi}=QFT\ket{\phi}$.
From their $\ell_2$ distance measure we can bound the \textit{trace-distance} between the two states by employing the Fuchs-van-de-Graaf inequality \cite{fuchs1999cryptographic}, which is tight for pure states \cite{kliesch2021theory}, when we restrict ourselves to states that amplitude-encode probability distributions, see App. \ref{appendix:trace-distance}. From here on out, all norms are assumed to be $\ell_2$ unless stated otherwise.

We aim to upload a target quantum state $\ket{\psi}$ into $n+m$ qubits by using QFT interpolation on state $\Tilde{\ket{\psi}}$, originally uploaded to $n$ qubits, which will be referred to as an \textit{$n$-qubit band-limited} state. An $n$-qubit band-limited state refers to a quantum state with any number of qubits that only has $2^n$ non-zero Fourier components. More precisely, a state with initial distribution $P_n$ discretized over a span $\Delta x_n$, will only have non-zero Fourier components within the band $2^n/\Delta x_n$.

Analogous to the Nyquist-Shannon sampling theorem \cite{nyquist1928certain, shannon1949communication}, if the target state $\ket{\psi}$ is $n$-qubit band-limited, the QFT interpolation technique is able to capture all the information of the underlying distribution, and the interpolation is perfect. However, that will not always be the case. Then, the best possible initial state $\ket{\psi}$ to interpolate from will be the one that minimizes $\lnorm{\ket{\psi}-\Tilde{\ket{\psi}}}=\lnorm{\ket{\Psi}-\Tilde{\ket{\Psi}}}$. Classically, the best band-limited approximation is realized by a convolution of the original signal with the sinc function \cite{whittaker1915xviii}, analogous to applying a low-pass filter to the signal, so that the high frequencies are cut off while maintaining the band-limited frequencies intact. This, however, is not as straightforward with quantum states, as they have to maintain their $\ell_2$ norm throughout the process, and a low-pass filter is not a unitary operation. Therefore, see App. \ref{appendix:sinc} for the detailed derivation, the optimal distance between states will be
\begin{equation}\label{eq:endbl}
\begin{split}
    \lnorm{\ket{\Psi}-\Tilde{\ket{\Psi}}}^2&=\frac{2\lnorm{\ket{\Psi_{out}}}^2}{1+\sqrt{1-\lnorm{\ket{\Psi_{out}}}^2}},\\
    &\leq2\lnorm{\ket{\Psi_{out}}}^2,
\end{split}
\end{equation}
where $\ket{\Psi_{out}}$ denotes the Fourier components of target state $\ket{\psi}$ outside of the $n$-qubit band-limit. The distance between the target state and the best possible interpolation will depend on the $\ell_2$ norm of the Fourier components that are filtered out. Using this result, we can bound the \textit{trace-distance} between the target and interpolated state by
\begin{equation}\label{eq:trace-dist}
\begin{split}
    \text{dist}_{\text{Tr}}(\ket{\psi}, \Tilde{\ket{\psi}})&\leq\sqrt{2}\lnorm{\ket{\Psi_{out}}}\sqrt{1-\frac{\lnorm{\ket{\Psi_{out}}}^2}{2}},\\
    &\leq\sqrt{2}\lnorm{\ket{\Psi_{out}}}.
\end{split}
\end{equation}

That being said, many applications do not have access to the the filtered distribution, as only the low-resolution values of the original distribution are available. The interpolated state will then suffer from aliasing effects, where the Fourier components of subsampled distributions are mixed with its own high-frequency modes due to the periodic nature of the discrete Fourier transform. In other words, the frequency components outside the $n$-qubit band limit, $\ket{\Psi_{out}}$, are added to the low frequency components, which we will call $\ket{\Psi_{in}}$, creating artifacts in the interpolated distribution.

The accuracy of the QFT interpolation algorithm will be worse in this approach, but we can still provide analytical bounds on the distinguishability due to the effects of aliasing. Now the distance between the target and interpolated state will be, refer to App. \ref{appendix:aliasing} for details,
\begin{equation}\label{eq:endal}
\begin{split}
    \lnorm{\ket{\Psi}-\Tilde{\ket{\Psi}}}^2&=\frac{2\lnorm{\ket{\Psi_{out}}}^2}{N}-\frac{(N-1)^2}{N}\\
    &\leq2\lnorm{\ket{\Psi_{out}}}^2,
\end{split}
\end{equation}
where $N\ge1$ is the normalization factor of the target state $\Tilde{\ket{\Psi}}$ due to the effects of aliasing.
Therefore, the upper bound on the \textit{trace-distance} under aliasing effects remains the same as Eq. \ref{eq:trace-dist}.

We have shown how the QFT interpolation algorithm approaches the target quantum state for a probability distribution. 
The smaller the norm of the high-frequency components, the better the interpolation will be. When the distribution is $n$-qubit band-limited, that is $\lnorm{\ket{\Psi_{out}}}=0$, it captures all the information of the underlying function and can be interpolated with as many qubits as needed. Even for non band-limited distributions, we show in Fig. \ref{fig:fig1} that with an initial uploading of 4 qubits, a Gaussian distribution can be interpolated with high fidelity to an exponentially larger space.

\section{Efficient QCT resampling of natural images}\label{sec:qct}

Interpolation methods, referred to also as resampling, are very common techniques in image processing. Moreover, natural images tend to have suppressed high-frequency components, and in particular, algorithms designed to process natural images tend to utilize the DCT \cite{ahmed1974discrete}. This real-to-real variant of the Fourier Transform is particularly suited for this type of signals. So much so, that some of the most prominent image processing techniques, like the image compression scheme JPEG \cite{wallace1992jpeg}, use this transformation as the basis of the algorithm.
The field of quantum image processing has experienced a lot of progress during the recent years \cite{latorre2005image, le2011frqi, zhang2013neqr, zhang2015qsobel, yao2017edge} due to the efficiency of such transformations in quantum computing. 
Therefore, resampling quantum encoded images using a QCT interpolation algorithm seems a natural step forward to generalize the technique to more dimensions.

\begin{figure*}[t]
    \centering
    \includegraphics[width=.34\textwidth]{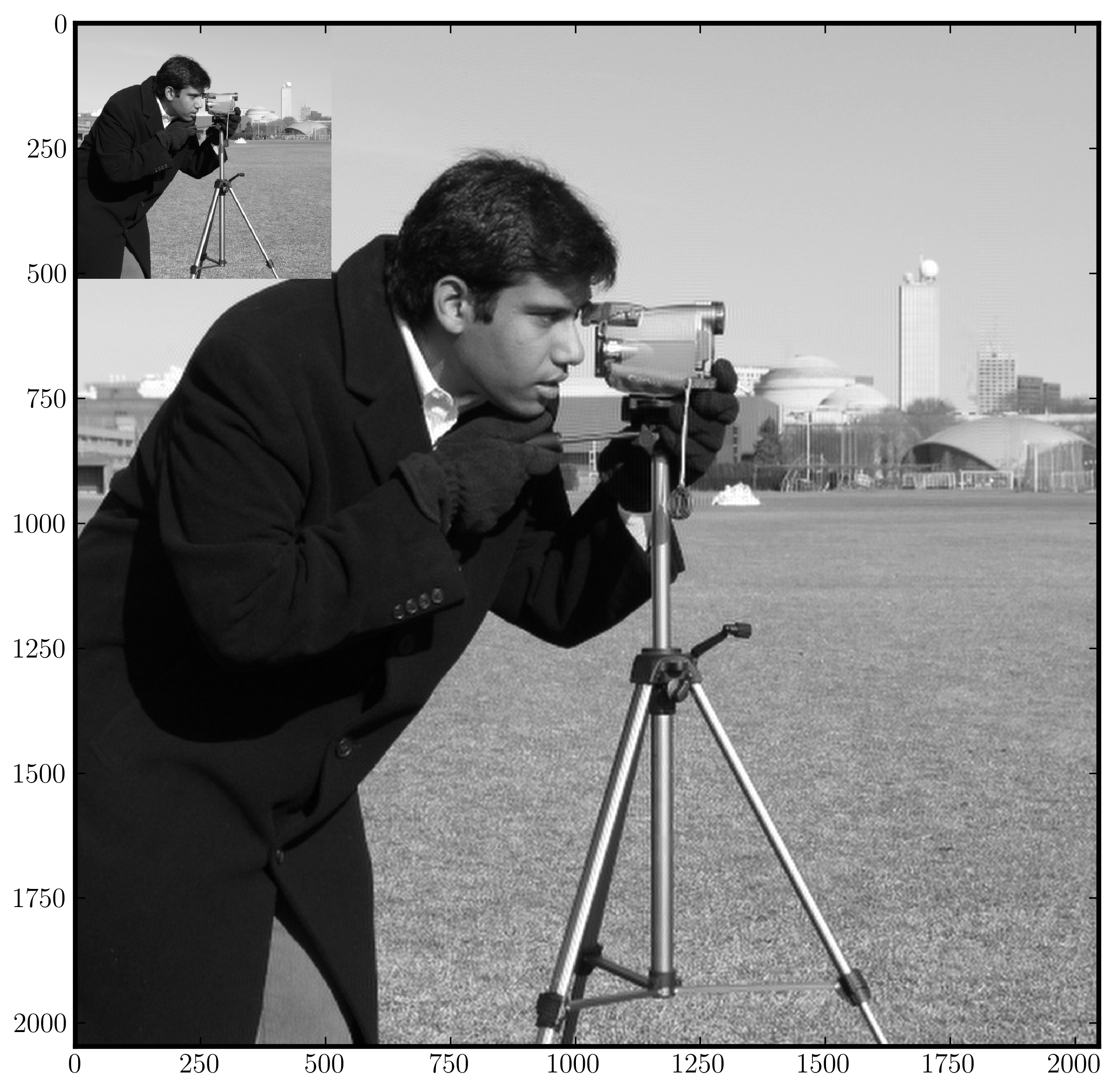}
    \quad\quad\quad\quad\quad\quad
    \includegraphics[width=.34\textwidth]{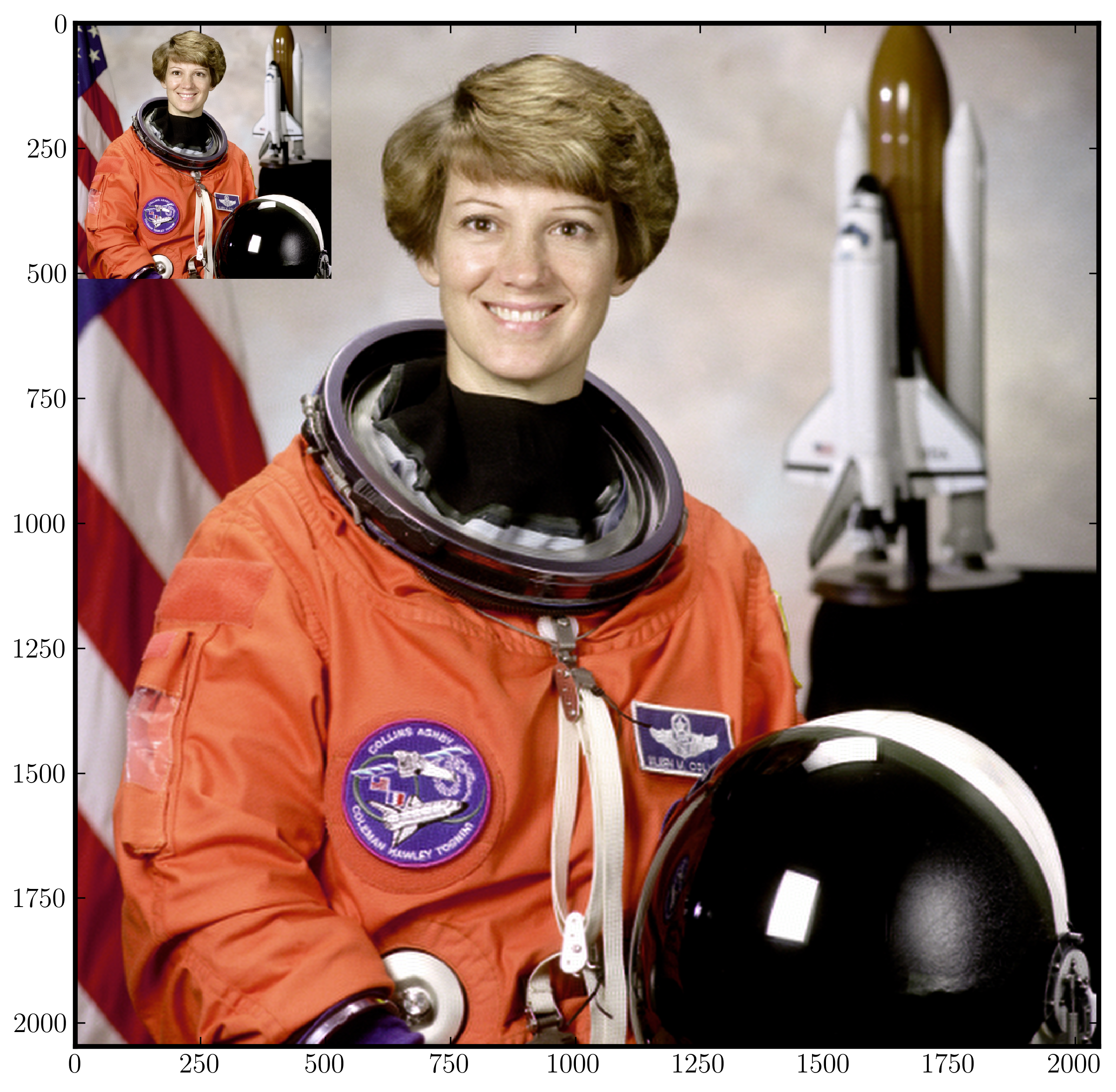}
    \caption{Example of the QCT resampling technique on images. (left) Grayscale image enlarged four times using the two dimensional QCT resampling technique. (right) RGB image enlarged four times using the two dimensional QCT resampling technique, ancillary qubits are used to label each color layer. Both quantum interpolation circuits require the same gate complexity regardless of the layers of the image.}
    \label{fig:interpolation-image}
\end{figure*}

Encoding images into a quantum state is a costly task. Most algorithms that attempt to upload a classical image into a quantum state \cite{le2011frqi, zhang2013neqr} inevitably scale with the pixel count of the image. This encoding techniques quickly become prohibitive for large images need to be uploaded. The proposed algorithm can alleviate part of this complexity, by applying QCT interpolation to resample an image in an efficient way after an initial uploading. This method also supports other uploading methods that rely on machine learning or quantum sensing techniques.

An efficient quantum implementation of the DCT is needed. In particular, we are interested in the type-II DCT, defined as
\begin{equation}
    \text{DCT-II}:\, X_k = \alpha_k \sum_{i=0}^{I-1}x_i\cos\left(\frac{(2i+1)k\pi}{2I}\right),
\end{equation}
where $\alpha_k$ are normalization constants, for $k=0,\ldots I-1$. 

\begin{figure}[h]
    \centering
    \resizebox{0.9\linewidth}{!}{\input{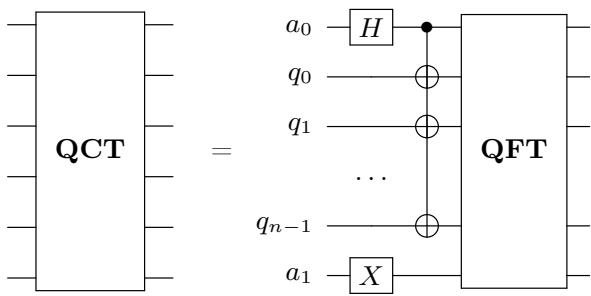}}
    \caption{Circuit example that implements a Quantum Cosine Transform (QCT) on the quantum state encoded in register $q=q_0,\ldots q_{n-1}$ using two ancilla qubits.}
    \label{fig:qct}
\end{figure}

There has been some work for quantum circuits that implement the DCT \cite{klappenecker2001discrete, pang2006quantum}, however, none are suitable for this implementation. Therefore, we need introduce a novel method of implementing the type-II DCT efficiently on a quantum computer. There is a direct mapping between the DCT-II and a DFT on a larger, symmetric space. Specifically, one needs to upload the original (real) signal $x_i$ into a $4I$ space in the following way. The signal $x_i$ is uploaded in the odd-indexed elements of the state in the first $2I$ entries. The inputs from $2I$ to $4I$ will mirror the first half. This can be mapped to direct operations on a quantum computer. Two ancilla qubits are added to the system, one as the most significant qubit, the other as the least significant one. A Hadamard gate on the first ancilla, followed by CNOT gates from that qubit to the rest of the signal-encoding qubits will copy the initial signal in a symmetric way over a $2I$ space. The second ancilla qubit only requires a single X gate, in order to encode the symmetric signal on the odd numbered elements of the quantum state. A QFT applied on the $n+2$ qubit system will yield the desired transformation, up to an overall scale factor. This transformation needs only to be inverted before the data is recovered. A circuit implementing the QCT is illustrated in Fig.~\ref{fig:qct}.

The QCT interpolation algorithm, understood as substituting the QFT by the QCT in the scheme presented in Sec. \ref{sec:qft}, can be extended to two, and up to any, dimensions. The image data that we want to process requires the following amplitude encoding. For two dimensions, values labelled by $\{x, y\}$ will be encoded into the amplitude of the quantum state $\ket{x}\ket{y}$ in the computational basis of registers $q_x$ and $q_y$. If this encoding is realized, applying the QCT circuit to both $q_x$ and $q_y$ registers separately (and simultaneously) achieves the 2-dimensional Discrete Cosine Transform over the data.

\textbf{Grayscale images.} The quantum resources needed to encode and interpolate a gray-scale image with $2^n\times2^n$ pixels using the proposed technique is $2n$ qubits, plus constant ancilla depending on the uploading method, $2m$ ancillary qubits used to enlarge the image to $2^{n+m}\times2^{n+m}$ pixels and 2 ancilla qubits for each dimension the QCT needs to be applied to. An example of an interpolated image using this method is displayed on Fig. \ref{fig:interpolation-image} (left). A $512\times512$ image (top left) has been enlarged four times in both axis. This implementation requires $2\times9+2\times2+4=26$ qubits, omitting encoding ancillas, to transform the original image to the enlarged space and has been simulated using the quantum simulation library \texttt{Qibo} using images available in the image processing library \texttt{scikit-image} \cite{scikit-image}.

\textbf{Multi-layer images.} We highlight a genuinely quantum speed-up that arises when working with multi-layer data. By properly encoding the multiple layers, we can process all the quantum data with a single call to the QCT interpolation algorithm. Starting from the same encoding technique used for a single image, each layer $l_i$ of the image will be labelled by state $\ket{l_i}$ of a new label quantum register $q_l$, with $\lceil\log(l)\rceil$ qubits, where $l$ is the total number of layers. That is, the value of pixel $\{x, y\}$ of layer $l_i$ will be encoded in the amplitude of the quantum state $\ket{x}\ket{y}\ket{l_i}$.

In this encoding, applying the QCT interpolation algorithm in the same way as one would for gray-scale images, acting on quantum registers $q_x$ and $q_y$ only, will perform the interpolation to all layers of the image in superposition, at no extra quantum cost. Since every pixel value is now entangled with its label, the amplitude interference that makes the QFT possible (and efficient) will only act on the amplitudes of pixel states that share the same label state. We showcase an example in Fig. \ref{fig:interpolation-image} (right), where we interpolate an RGB image using two qubits to keep track of the color channel. Since the pixel dimensions are the same as Fig. \ref{fig:interpolation-image} (left), the depth of the quantum circuit required for interpolation is the same for both instances.

The speed-up provided by acting on all layers in superposition is made more apparent the more layers, or data instances sharing the same shape, that are encoded in the proposed way. Additionally, further quantum advantage can be achieved when implementing quantum transformations to all subsets within an image at the same time. Beyond that, this can be extended to other types of data uploaded in superposition. If the initial cost of uploading data using a label register is overcome, any quantum processing using a transformation such as the QFT or QCT can be applied to all members of the superposition in singular cost. However, we need to keep in mind that if the data in superposition needs to then be extracted via measurements, the advantage that the parallel processing power introduces will be lost.

\begin{figure*}[t]
    \centering
    \includegraphics[width=\linewidth]{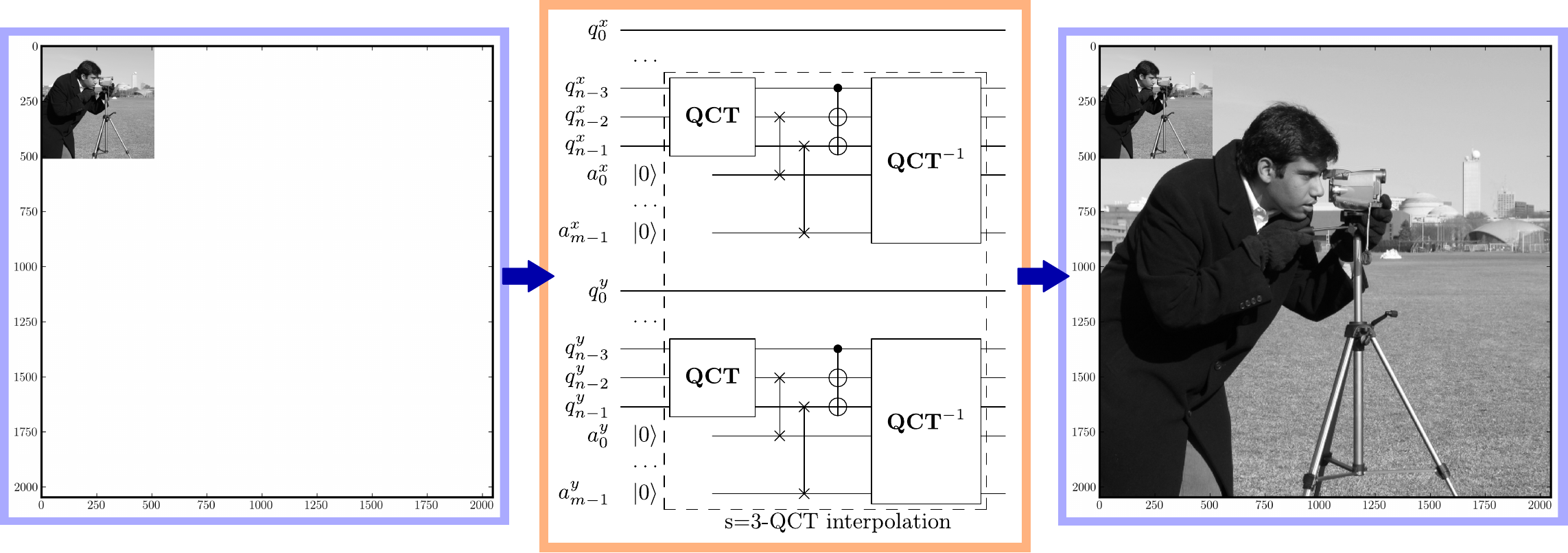}
    \put(-508,163){\textbf{a})}
    \put(-331,163){\textbf{b})}
    \put(-164,163){\textbf{c})}
    \caption{Interpolation of a grayscale image using the JPEG-inspired, s=3-QCT method. a) Depicts the original image embedded in the larger subspace. b) Showcases the quantum circuit that performs s=3-QCT interpolation for two dimensional data. The decomposition of the QCT blocks are shown in Fig.~\ref{fig:qct}, and the two ancilla needed are implied. The interpolation algorithm for two dimensions can be applied simultaneously, therefore not incurring extra depth cost. The algorithm is independent on the original system size, as it only acts on the subspace $s$ and the added ancilla $m$. c) Shows the result of the interpolated image using $m=2$ ancilla for each of the two dimensions. The original image is shown for scale.}
    \label{fig:qjpeg}
\end{figure*}

\section{Subspace-QCT resampling of natural images}\label{sec:qjpeg}

The power of parallel QFT computation can be extended beyond actions on different sets of data encoded in superposition. Subsets of a single data entry can also be processed in parallel.

Correlations present in natural data tend to be localized within short distances. Indeed, natural images are usually built from large structures (in number of pixels) that do not necessarily correlate with the rest of the picture. The JPEG protocol for image compression \cite{wallace1992jpeg} exploits this fact by performing DCT onto $8\times 8$ blocks of the image. This way, the small DCT only capture short range correlations, and the following steps of the protocol can be described using $8\times 8$ matrices.

A similar procedure can be applied in a quantum circuit. All the $8\times 8$ pixel matrices of the image encoded in the proposed way are stored in the superposition of the 3 least significant qubits. The rest of the system can be understood in the same way we considered the label register for multi-layer images. Applying a transformation to the 3 least significant qubits, a QCT in this case, achieves in constant depth the transformation of all subspaces of the image.

To be precise, this JPEG-inspired interpolation algorithm using the DCT proceeds as follows. Given an image encoded into a quantum state in registers $\ket{x}\ket{y}\ket{l}$, perform the QCT on the last 3 qubits of registers $\ket{x}$, and simultaneously in the last 3 qubits of register $\ket{y}$. Introduce the ancilla qubits needed for the interpolation after the third least significant qubit on both registers. Undo the QCT transformation on the extended space. This achieves image interpolation at a depth constant with the original system size. Given a fixed subspace $2^s$, $s=3$ in the JPEG-inspired procedure, this algorithm resamples an $2^n$ signal into a $2^{n+m}$ space in complexity $\order{(s+m)^2}$, the complexity of the algorithm no longer depends on the size of the original image. A circuit depicting this algorithm is sketched in Fig.~\ref{fig:qjpeg}b. 

This algorithm can be generalized to any subspace $s$, and can introduce improvements depending on the underlying structure of the signal. In particular, an \textbf{s-QCT} with $s=n$ is equivalent to the full QCT approach presented in the previous section.  

We compare the performance of the presented algorithms for image interpolation. A trial image is first down-scaled using the recommended classical algorithm as provided by the OpenCV library \cite{opencv_library}. Then, the down-scaled image is interpolated back into its original size, and compared to the base image. Two metrics are commonly used in order to assess the accuracy of image interpolation \cite{hore2010image}, the Peak Signal-to-Noise Ratio (PSNR) and Structural SIMilarity (SSIM) \cite{wang2004ssim}, detailed in App. \ref{appendix:metrics}. The interpolation is performed using 4 algorithms, classical bicubic interpolation, QFT interpolation, QCT interpolation and s=3-QCT interpolation.

In Table \ref{tab:comparison} we compare the two metrics for the different interpolation methods. Bicubic interpolation is one of the go-to methods for classical image interpolation, it uses the adjacent $4\times 4$ pixels to compute each new value. This method achieves the best results by a slight margin when compared to the better of the quantum algorithms presented. However, it incurs a computational cost of $\order{N_p^2}$ \cite{acharya2007bicubiccomplex}, where $N_p$ is the number of pixels in the image, exponentially more expensive than the quantum counterparts. Alternatively, the best performing quantum method is the s=3-QCT, which is also the least expensive in terms of computational complexity. By focusing on small subsets of the image, the algorithm is both faster, due to processing data in superposition, and more accurate in reproducing the original data.

\begin{table}[t]
    \centering
    \begin{tabular}{|p{1cm}||p{1.66cm}|p{1.66cm}|p{1.66cm}|p{1.66cm}|}
         \hline
         m=1 & Bicubic & QFT & n-QCT & 3-QCT \\\hline
         PSNR &  30.095 & 27.395 & 29.930 & 29.988 \\\hline
         SSIM & 0.880 & 0.829 & 0.871 & 0.878 \\\hline
    \end{tabular}
    \caption{Comparison of the Peak Signal-to-Noise Ratio (PSNR) and Structural SIMilarity (SSIM) of the gray-scale \textit{camera} image, after different methods of interpolation. The image is downsized to half its original size ($m=1$) via a classical algorithm using the pixel area relation of the image implemented in OpenCV. The image is interpolated to the original size and compared to the base image. The classical bicubic inteprolation scheme yield the best PSNR and SSIM values, closely followed by the s=3-QCT. From the quantum algorithms, the best choice is the JPEG-inspired s=3-QCT method, it provides the best results and is the most efficient option.}
    \label{tab:comparison}
\end{table}

By using techniques of classical image processing we have enhanced the quantum interpolation algorithm, both in terms of the accuracy of the reconstruction and complexity. The ideas presented in the previous sections about computing in superposition have been extended to small subsets of the same image. By fixing the subspace where the QCT acts on to 3 qubits, the interpolation algorithm no longer scales with the size of the original image, allowing for a very efficient circuit for interpolation of natural data in a quantum computer.


\section{Conclusion} 

We have proposed efficient quantum techniques to interpolate distributions, all exploiting the Quantum Fourier Transform. We show that this QFT interpolation techniques achieves favourable results on data with negligible high-frequency components, as well as in practical examples using natural images.

This techniques can be used to enhance current uploading algorithms by focusing on smaller scale and accurate uploading techniques that can then be efficiently resampled via interpolation algorithms. It can also be extended to any uploading technique that deals with non-classical band-limited quantum signals. Additionally, this interpolation technique can be used on genuinely quantum states, but the interpolation will be extended along computational basis states which might fail to capture correlations that go beyond that.

We have also showcased the power of processing quantum data in superposition. 
Encoding different sets of data using label ancillas allows for the implementation of quantum transformations in parallel. Furthermore, by extending native ideas of natural image processing, used in the JPEG procedure for image compression, we can employ this parallel processing power of a quantum circuit to gain a substantial advantage. By processing subspaces of an image simultaneously using quantum superposition, we can perform interpolation in a complexity that is constant with the original size of the image.

We would like to further highlight the implementation of efficient quantum transformations in order to provide quantum advantages to algorithms that might not initially rely on them. After all, at the core of the efficiency of Shor's factorization algorithm~\cite{shor1999polynomial} is the use of the QFT. Looking into areas where these types of transformations are extensively used may result in further avenues for quantum advantage. Natural image and signal processing are fields that have evolved around such transformations, further work might borrow from such well understood fields in order to enhance quantum algorithms in different ways or explore interpolation techniques with even more advanced transformations. Compression techniques that rely on these transformations might also provide quantum advantages.

The code used to simulate the quantum circuits presented is available online \cite{github}.

\section*{Acknowledgements}

The author would like to acknowledge J. I. Latorre and I. Roth for fruitful discussions. 

\bibliography{Citations}

\appendix

\clearpage

\section{Enhancing a unary uploading}\label{appendix:unary}

We illustrate how the QFT resampling algorithm can enhance uploading techniques using the amplitude distribution in the unary basis presented in Ref. \cite{ramos2021quantum}. This proposal requires only nearest-neighbor connectivity and employs gates that are well suited for implementation on near-term quantum devices. The distribution is uploaded in the unary basis where only one qubit is in state $\ket{1}$ while the others remain at $\ket{0}$, ie. $\ket{10000}, \ket{01000}, \ldots \ket{00001}$. This, however, reduces the Hilbert space available for computation and requires linear depth with the number of amplitudes needed to upload, limiting its usability. While this basis helps with gate application and control of the device, one would want to exploit the exponential Hilbert space that qubits support. 

\begin{figure}[b]
    \centering
    \resizebox{0.9\linewidth}{!}{\input{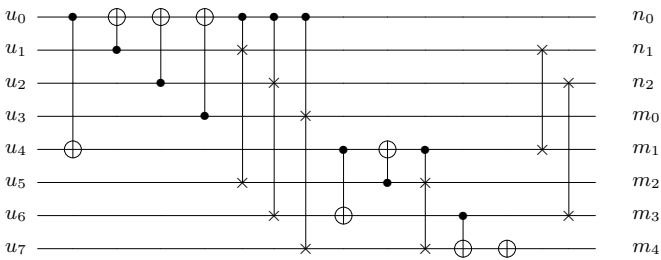}}
    \caption{Circuit example that encodes an $8-\text{to}-3$ basis change from the unary basis to the binary one. The $u$ register depicts the unary register where $u_0$ is the most significant qubit in unary representation. On the left-hand side the labels for the binary qubits $n$, and the ancilla qubits $m$, are depicted. The final SWAP gates can be neglected if we keep track of the qubit order and perform classical reshuffling.}
    \label{fig:un2bin}
\end{figure}

\begin{figure}[t!]
    \centering
    \includegraphics[width=.8\linewidth]{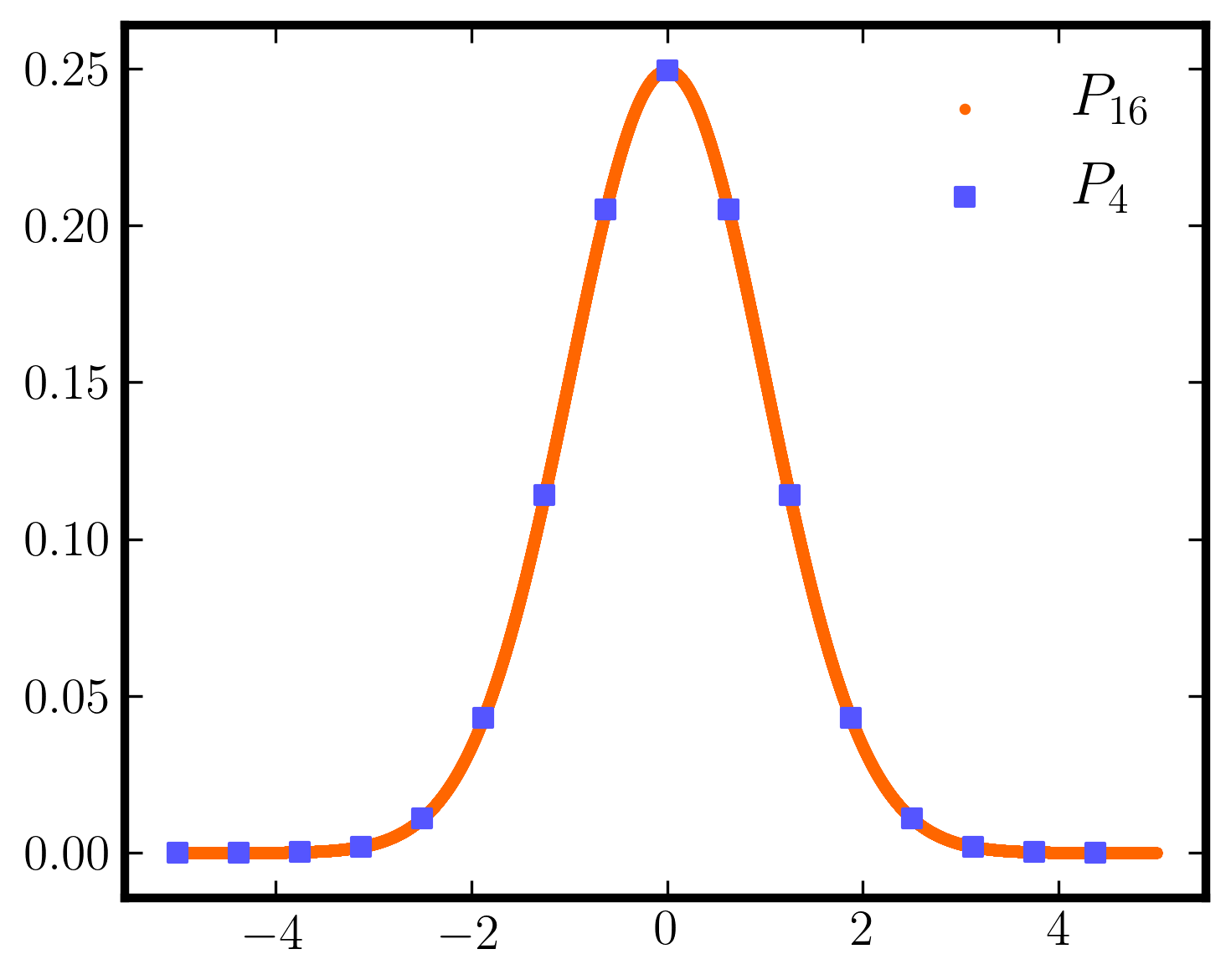}
    \caption{Comparison of a Gaussian probability distribution uploaded using the unary basis to $16$ qubits $P_4$ that is enlarged into $P_{16}$ using the QFT interpolation scheme to span the full $2^{16}$ Hilbert space.}
    \label{fig:interpolation-unary}
\end{figure}

In order to reclaim the lost Hilbert space using QFT interpolation we propose the following. Upload a probability distribution into the unary basis over a $2^n$ qubit register. Then, perform a unitary change of basis from unary to binary basis \footnote{brought to my attention in a post by Craig Gidney in \texttt{https://quantumcomputing.stackexchange.com
/questions/5526/garbage-free-reversible-
binary-to-unary-decoder-construction}}, as detailed in Alg. \ref{alg:un2bin} and illustrated with a small example in Fig. \ref{fig:un2bin} for $n=3$. After the change of basis, the quantum state contains $n$ qubits storing the superposition and $2^n-n$ clean ancillas at state $\ket{0}$. This allows the implementation of the QFT interpolation algorithm using the clean ancilla register to encode the high frequency components. Now the full extent of the available Hilbert space is used to encode the interpolated probability distribution. Shown in Fig. \ref{fig:interpolation-unary} are the simulation results of a $16$ qubit total QFT interpolation with unary uploading.
\begin{algorithm}[t!]
\caption{\label{alg:un2bin} Unary to binary encoding.}
\DontPrintSemicolon
\SetKwFunction{FMain}{Unary2Binary}
\SetKwProg{Fn}{Function}{:}
\Fn{\FMain{$n$}}
{\;
    $c \leftarrow {\rm Circuit}(2^n)$\;
    Ensure $q = 0 $ \;
    \For{$i \leftarrow 0 \;\KwTo \; n-1$}
    {
        $qq \leftarrow 2^{n-i-1}$\;
        $c$.add(CNOT($q,\; qq$)) \;
        \For{$j \leftarrow 1\; \KwTo \; qq-1$}
        {
        $c$.add(CNOT($q+j,\; q$))\;
        }
        \For{$j \leftarrow 1\; \KwTo \; qq-1$}
        {
        $c$.add(SWAP($q+j,\; q+j+qq$).controlled\_by($q$))\;
        }
        $q \leftarrow q+qq$\;
    }
    $c$.add(X($2^n-1$))\;
    \KwRet{${\rm Binary\;basis\;in\;qubits\;}\{2^n-2^i\}\; {\rm with}\;i \leftarrow n\; \KwTo \; 1$} }
\end{algorithm}

\section{Trace-distance for pure states encoding real probability distributions}\label{appendix:trace-distance}

The trace-distance for pure quantum states is defined as
\begin{equation}
    \text{dist}_{\text{Tr}}(\ket{\psi},\ket{\phi})=\frac{1}{2}\lonorm{\ket{\psi}\bra{\psi}-\ket{\phi}\bra{\phi}}.
\end{equation}
At the same time, for pure quantum states, the Fuchs-van-de-Graaf inequality is tight, meaning that
\begin{equation}\label{eq:fvdg}
    \lonorm{\ket{\psi}\bra{\psi}-\ket{\phi}\bra{\phi}}=2\sqrt{1-\abs{\bra{\psi}\ket{\phi}}^2}.
\end{equation}
Recall then, that the $\ell_2$ norm of the same pure states can be decomposed as
\begin{equation}
    \lnorm{\ket{\psi}-\ket{\phi}}^2=\lnorm{\ket{\psi}}^2+\lnorm{\ket{\phi}}^2-2\,\text{Re}\left(\bra{\psi}\ket{\phi}\right).
\end{equation}
In particular, when the pure states being considered are amplitude encodings of probability distributions, that is, all amplitudes are real valued, the $\ell_2$ norm can be simplified as 
\begin{equation}
    \lnorm{\ket{\psi}-\ket{\phi}}^2=2(1-\abs{\bra{\psi}\ket{\phi}}).
\end{equation}
Therefore one can substitute the value
\begin{equation}
    \abs{\bra{\psi}\ket{\phi}}=\left(1-\frac{\lnorm{\ket{\psi}-\ket{\phi}}^2}{2}\right)
\end{equation}
in Eq. \ref{eq:fvdg} in order to recover the trace-distance from the $\ell_2$ distance in this particular case. Precisely,
\begin{equation}
    \text{dist}_{\text{Tr}}(\ket{\psi},\ket{\phi})=\lnorm{\ket{\psi}-\ket{\phi}}\sqrt{1-\frac{\lnorm{\ket{\psi}-\ket{\phi}}^2}{4}}.
\end{equation}

\section{Distance for \textit{band-filtered} pure states}\label{appendix:sinc}

The \textit{bra-ket} notation will be dropped in the derivation section for clarity.

The band-limited version of target state $\Psi$ can be defined as 
\begin{equation}
    \Tilde{\Psi}=\frac{\Psi_{in}}{\lnorm{\Psi_{in}}},
\end{equation}
where $\Psi_{in}$ are the components of the target $\Psi$ that fall within the $n$-qubit band-limited space. Alternatively, $\Psi_{out}$ are the components outside of the $n$-qubit band-limit, and $\Psi_{in}+\Psi_{out}=\Psi$.

Therefore, the difference between the target state and the initial state used in the interpolation is
\begin{equation}
    \lnorm{\Psi-\Tilde{\Psi}}^2=\lnorm{\Psi_{out}}^2+\left(1-\frac{1}{\lnorm{\Psi_{in}}}\right)^2\lnorm{\Psi_{in}}^2.
\end{equation}
Developing the square and using the relation $\lnorm{\Psi_{in}}^2+\lnorm{\Psi_{out}}^2=1$ results in
\begin{equation}
    \lnorm{\Psi-\Tilde{\Psi}}^2=2+2\frac{\lnorm{\Psi_{out}}^2-1}{\sqrt{1-\lnorm{\Psi_{out}}^2}}
\end{equation}
that will only depend on the norm of the high frequency components. By combining the terms into
\begin{equation}
    \lnorm{\Psi-\Tilde{\Psi}}^2=2\frac{\sqrt{1-\lnorm{\Psi_{out}}^2}-1+\lnorm{\Psi_{out}}^2}{\sqrt{1-\lnorm{\Psi_{out}}^2}}
\end{equation}
and multiplying both terms in the fraction by $\sqrt{1-\ltnorm{\Psi_{out}}^2}+1$ we reach the simplified form shown in the main text,
\begin{equation}
    \lnorm{\Psi-\Tilde{\Psi}}^2=\frac{2\lnorm{\Psi_{out}}^2}{\left(1+\sqrt{1-\lnorm{\Psi_{out}}^2}\right)}.
\end{equation}

\section{Distance for aliased pure states}\label{appendix:aliasing}

The derivation of the $\ell_2$ distance between the target and aliased quantum state will proceed similarly to the band-limited one, with more detail on the normalization constant $N$.

The difference now can be written as
\begin{equation}
    \lnorm{\Psi-\Tilde{\Psi}}^2=\lnorm{\Psi-\frac{1}{N}\left(\Psi_{in}+\Phi\right)}^2,
\end{equation}
where $N=\lnorm{\Psi_{in}+\Phi}$ is the normalization factor of the interpolated state $\Tilde{\Psi}$ and $\Phi$ is the aliasing effect of the off band-limit components which satisfies $\lnorm{\Phi}=\lnorm{\Psi_{out}}$.  Notice that $N\ge1$, since 
\begin{equation}
    \lnorm{\Psi_{in}+\Phi}^2\ge\lnorm{\Psi_{in}}^2+\lnorm{\Phi}^2=\lnorm{\Psi_{in}}^2+\lnorm{\Psi_{out}}^2=1,
\end{equation}
due to $\Psi_{in}$ and $\Psi_{out}$ composing to reconstruct $\Psi$, which is a normalized quantum state.

The norm outside the band limit remains the same, leaving
\begin{equation}
    \lnorm{\Psi-\Tilde{\Psi}}^2=\lnorm{\Psi_{out}}^2+\lnorm{\left(1-\frac{1}{N}\right)\Psi_{out}-\frac{1}{N}\Phi}^2.
\end{equation}
The second term can be decomposed, making the distance then
\begin{equation}\label{eq:aliasing1}
\begin{split}
    \lnorm{\Psi-\Tilde{\Psi}}^2=&\lnorm{\Psi_{out}}^2+\left(1-\frac{1}{N}\right)^2\lnorm{\Psi_{in}}^2+\frac{1}{N^2}\lnorm{\Phi}^2\\
    &-2\left(1-\frac{1}{N}\right)\frac{1}{N}\,\text{Re}(\bra{\Psi_{in}}\ket{\Phi}).
\end{split}
\end{equation}
In order to get rid of the last term, we recall that the norm $N$ of the aliased state is
\begin{equation}
    N^2=\lnorm{\Psi_{in}+\Phi}^2=\lnorm{\Psi_{in}}^2+\lnorm{\Phi}^2+2\,\text{Re}(\bra{\Psi_{in}}\ket{\Phi}),
\end{equation}
and since $\lnorm{\Psi_{in}}^2+\lnorm{\Phi}^2=\lnorm{\Psi_{in}}^2+\lnorm{\Psi_{out}}^2=1$, we can substitute
\begin{equation}
    2\,\text{Re}(\bra{\Psi_{in}}\ket{\Phi})=N^2-1
\end{equation}
into Eq. \ref{eq:aliasing1} to reach
\begin{equation}
\begin{split}
    \lnorm{\Psi-\Tilde{\Psi}}^2=&\lnorm{\Psi_{out}}^2+\left(1-\frac{1}{N}\right)^2(1-\lnorm{\Psi_{out}}^2)\\
    &+\frac{1}{N^2}\lnorm{\Psi_{out}}^2-\left(1-\frac{1}{N}\right)(N^2-1)\frac{1}{N}.
\end{split}
\end{equation}
This equation simplifies to the form shown in the main text,
\begin{equation}
    \lnorm{\Psi-\Tilde{\Psi}}^2=\frac{2\lnorm{\Psi_{out}}^2}{N}-\frac{(N-1)^2}{N}.
\end{equation}

\section{PSNR and SSIM for image interpolation}\label{appendix:metrics}

The PSNR between the original image $f$ and the interpolation result $g$ is defined as 
\begin{equation}
    PSNR(f, g) = 10\log_{10}\left(255^2/MSE(f,g)\right),
\end{equation}
with the Mean Squared Error (MSE) being
\begin{equation}
    MSE(f, g) = \frac{1}{NM}\sum_{i=1}^M\sum_{j=1}^N\left(f_{ij}-g_{ij}\right)^2
\end{equation}
for $M\times N$ images.
The higher the PSNR value, the closer the images are in terms of numerical value. Generally, one aims at a PSNR of 30 or above. 

The SSIM index is another image quality metric that is correlated with how humans perceive images. It compares three metrics, the luminance $l$, the contrast $c$ and the structure $s$ of two images $f$ and $g$. Each element is defined as 
\begin{equation}
    l(f, g) = \frac{2\mu_f\mu_g+c_1}{\mu_f^2+\mu_g^2+c_1},
\end{equation}
\begin{equation}
    c(f, g) = \frac{2\sigma_f\sigma_g+c_2}{\sigma_f^2+\sigma_g^2+c_2},
\end{equation}
\begin{equation}
    s(f, g) = \frac{\sigma_{fg}+c_3}{\sigma_f\sigma_g+c_3},
\end{equation}
where $\mu_{f,g}$ is the average of the image, $\sigma_{f,g}^2$ is the variance, $\sigma_{fg}$ the covariance and $c_{1,2,3}$ are constants to avoid $0$ on the denominator. The SSIM is the product of the three metrics,
\begin{equation}
    SSIM(f, g) = l(f, g)\,c(f, g)\,s(f, g),
\end{equation}
and is equal to $1$ when the two images are the same.

\end{document}